\documentclass[pre,twocolumn,showpacs,preprintnumbers,amsmath,amssymb]{revtex4}
\usepackage{graphics}

\begin{document}

\title{Core reconstruction in pseudopotential calculations}
\author{J R Trail}\email{j.r.trail@bath.ac.uk}
\author{D M Bird}
\affiliation{Department of Physics, University of Bath, Bath BA2 7AY, UK}

\date{June 1999}

\begin{abstract}
A new method is presented for obtaining all-electron results from a
pseudopotential calculation.  This is achieved by carrying out a
localised calculation in the region of an atomic nucleus using the
embedding potential method of Inglesfield [J.Phys.\ C {\bf 14}, 3795
(1981)].  In this method the core region is \emph{reconstructed}, and
none of the simplifying approximations (such as spherical symmetry of
the charge density/potential or frozen core electrons) that previous
solutions to this problem have required are made.  The embedding
method requires an accurate real space Green function, and an analysis
of the errors introduced in constructing this from a set of numerical
eigenstates is given.  Results are presented for an all-electron
reconstruction of bulk aluminium, for both the charge density and the
density of states.
\end{abstract}


\maketitle

\section{Introduction}
Total energy pseudopotential methods have taken pride of place in the
first principles simulation of condensed matter in recent years due to
their efficient use of computing resources and their suitability for
structural optimisation \cite{payne92}.  However, the charge density
resulting from a pseudopotential calculation is incorrect in the
region near the atomic nuclei - it does not include core states, and
the valence states have the wrong structure.  In order to obtain
accurate values for any quantity that depends on the true charge
density, such as hyperfine interactions or X-ray structure factors, we
must obtain the correct electron charge density from the
pseudopotential calculation.  Other methods are available that
calculate the states of all the electrons in the system
(Full-potential Linear Augmented Plane Wave (FLAPW)\cite{hill98},
Projector Augmented Plane Wave (PAW)\cite{blochl94}, Linear Muffin Tin
Orbital (LMTO)\cite{turzhevsky94}, KKR Green function \cite{huhne98}
and Tight Binding \cite{goringe97}), but these methods tend to be more
computationally expensive, not as well suited to structural
optimisation, or are less accurate than pseudopotential methods.

In view of this it is desirable to extend the pseudopotential method
by adding an extra step after the pseudo-system has been solved, ie to
choose an atom for which we require the core and correct valence
states, and \emph{reconstruct} these correct states.  It would be
hoped that solving for one atom with different boundary conditions on
a sphere surrounding it would be fairly straightforward.  However, a
number of difficulties quickly present themselves.  The purpose of
this paper is to describe and validate a new procedure for carrying
out this \emph{core reconstruction} that makes essentially the same
\emph{physical} approximations as the FLAPW method, and so can be
expected to provide the same accuracy.

This problem has been addressed by several previous workers.  In their
paper Gardner and Holzworth \cite{gardner86} reconstruct the correct
states for isolated Si and Ru atoms from the pseudo-atom, by applying
direct integration, effectively `inverting' the pseudopotential.  They
obtain good results, showing that this reconstruction approach is at
least possible.  However, reconstructing the states of an atom in a
lattice is considerably more complex.  For this case the valence
states form a continuum, so the reconstruction must be fitted into the
band structure of the lattice, and in addition to this the potential
is not spherically symmetric.  Vack\'a\v{r} and \v{S}im\.unek
\cite{vackar94} describe a method for reconstructing the states for a
pseudo-atom within a lattice.  Their method relies on direct
integration and assumes the charge density, boundary conditions and
self-consistent potential are spherically symmetric, although the core
states are allowed to relax.  The errors in the resulting
eigenfunctions are fairly large, although they do obtain the correct
nodal form for the eigenstates.  Kuzmiak et al\cite{kuzmiak91} perform
a pseudopotential calculation, and orthogonalise the resultant
pseudo-states to the original core states.  This would work for the
original formulation of pseudopotential methods, where the
pseudopotential is defined in this way, but for modern norm-conserving
pseudopotentials this does not give the correct solution to the
problem and the errors present are difficult to control or even
quantify.  The most complete solution to the problem presented so far
is that due to Meyer et al \cite{meyer95}.  In their method the
correct states are reconstructed by direct integration.  In order to
decouple the radial wave equations for the reconstruction calculation
they make the assumption of spherical symmetry of the self consistent
potential, but asymmetric boundary conditions for the valence states
are allowed. Within their scheme the core is still frozen.

In this paper a new method for performing this kind of core
reconstruction is described that does not make any of the assumptions
of previous approaches.  The method presented here follows a different
path to achieving the reconstruction, does not require spherical
averaging of the self-consistent potential, provides an aspherical
charge density, and does not assume a frozen core.  The first step is
to carry out a plane-wave pseudopotential calculation for the system
of interest, and construct the single particle Green function for the
valence electrons present in this system.  This Green function is then
used to create an \emph{embedding potential}, as described by
Inglesfield \cite{inglesfield81}.  An all-electron localised atomic
calculation is then carried out in a space containing the core region
of the atom of interest, with the embedding potential taking into
account the rest of the atoms in the lattice.  Effectively the system
that is solved for is one all-electron atom in a lattice of
pseudo-atoms.  This approach preserves the full generality and
flexibility of the pseudopotential method.  In addition, the
all-electron calculation is carried out only for the atom(s) of
interest, so computational effort can be applied sparingly.

In the next section a brief description is given of the embedding
approach, how it is applied in this case, and how it relies on an
accurate knowledge of the Green function for the substrate system.
Appendix\ \ref{sec:appa} gives a more complete description of the
method and its properties.  In section \ref{sec:green} we describe how
the spectral representation can be efficiently applied in order to
construct an accurate Green function, and how the incompleteness of
the available set of states introduces a significant error that must
be corrected.  Combining the results of these two sections we then
obtain the embedding potential from the Green function and perform a
localised calculation using the embedding terms to take into account
the rest of the lattice, as described in section
\ref{sec:reconstruction}.  Section \ref{sec:results} gives the results
of a reconstruction carried out for bulk FCC aluminium, and the
convergence of the method is discussed.  Rydberg atomic units are used
throughout the paper.

\section{Embedding}
\label{sec:embedding}

\begin{figure}
\begin{center}
\includegraphics*{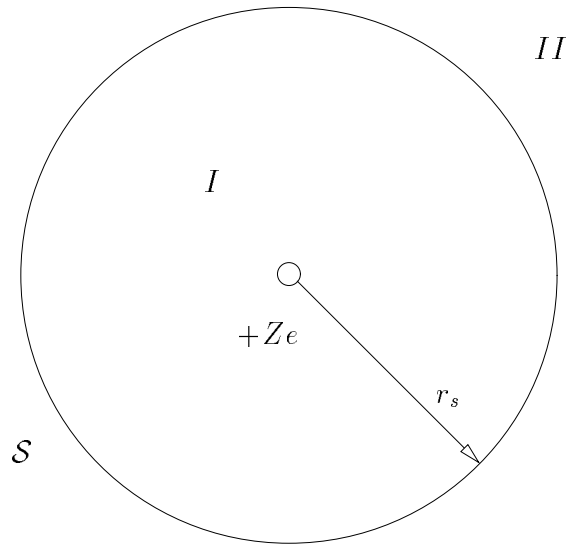}
\end{center}
\caption{Geometry of embedding calculation.  Region $II$ is the
substrate region, region $I$ is the embedded region, and the
all-electron states are reconstructed from a knowledge of the
pseudo-states on the surface $\mathcal S$.}
\label{fig1}
\end{figure}

An embedding method can be thought of as solving for a system in a
sub-domain of space, denoted region $I$ in what follows, where the
influence of the system outside of this sub-domain (region $II$) is
taken into account from a previous solution, and is \emph{not}
recalculated.  This is shown in Fig.\ \ref{fig1} for the core
reconstruction case considered here. The embedding surface, $\mathcal
S$, separates regions $I$ and $II$.  Region $I$ is a sphere of radius
$r_s$ enclosing the core region of the atom where the pseudo-states
are incorrect and region $II$ is the remainder of the lattice of
pseudo-atoms.  It is implicitly assumed throughout this work that
norm-conserving pseudopotentials are used, so that wavefunctions
between core regions are correct.  The embedding surface $\mathcal S$
is assumed to be in such a region; it follows that core regions do not
overlap, hence $r_s > r_c$ where $r_c$ is the largest pseudopotential
core radius.

Using Inglesfield's method \cite{inglesfield81} an `embedding
potential' is obtained from the substrate system, and added to the
Hamiltonian for the embedded region.  This embedding potential ensures
that the states of the system in the embedded region satisfy the
correct boundary conditions.  Inglesfield's method has the advantage
that it requires knowledge of the properties of the substrate system
only on the surface separating the embedded and substrate regions.  In
appendix\ \ref{sec:appa} a brief description of the derivation of the
embedding method is given, followed by two expressions for the
embedding potential in terms of the Green function for the substrate
system, one of which has not appeared in the literature before.

Within a continuum the Hamiltonian of the embedded system takes the
form
\begin{equation}
H_{emb}(E)= H_I+ \delta({\mathbf r}_{s}-{\mathbf r}) \left[
\frac{\partial}{\partial n_s} - \Gamma({\mathbf r}_{s},{\mathbf
r}'_{s};E) \right]
\label{eq1}
\end{equation}
where $H_{emb}(E)$ is the embedded Hamiltonian that yields the states
with correct boundary conditions, $H_I$ is the normal Hamiltonian for
the embedded region, and $\Gamma$ is the embedding potential.  It
should be understood at this point that $\Gamma({\mathbf
r}_{s},{\mathbf r}'_{s};E)$ acting on a function denotes the
integration over the surface $\mathcal S$, as described in appendix\
\ref{sec:appa}.  Equation (\ref{eq1}), when solved in region $I$, will
give the correct solution for the system represented by regions $I$
and $II$, with region $II$ represented entirely through the embedding
potential term.  The embedding potential required in equation (Eq.\
(\ref{eq1})) is given by the operator
\begin{equation}
\Gamma=-{\mathcal G}^{-1}.\left( {\mathcal I}-\frac{\partial {\mathcal
G}}{\partial n_s'} \right)
\label{eq2}
\end{equation}
where $\mathcal G$ is the matrix representation of the Green function
on the surface $\mathcal S$ in terms of a set of basis functions
orthonormal over the surface, and the derivative is the normal
derivative of $\mathcal G$ outward from the surface and with respect
to the second spatial variable of the Green function.  A derivation of
Eqs. (\ref{eq1}) and (\ref{eq2}) is given in appendix\ \ref{sec:appa}.

\section{Green Function and Embedding potential}
\label{sec:green}
The embedding method of the previous section is to be applied with the
substrate system represented by the results of a plane-wave
pseudopotential calculation, hence the Green function on the embedding
surface must be obtained from the Bloch states expanded in
plane-waves.  The natural way of constructing this Green function is
via the spectral representation, and a finding good approximation to
this spectral representation is the concern of this section.

For the periodic system the states are characterised by two quantum
numbers, the discrete band index $n$ and the continuous crystal
momentum $\mathbf k$, so the spectral representation takes the form
\begin{equation}
G({\mathbf r},{\mathbf r}';E)=\sum_n \int_{BZ} \frac{ \Psi_{\mathbf
k}^n({\mathbf r}) \Psi^{n*}_{\mathbf k}({\mathbf r}') }{ E_n({\mathbf
k})-E } d^3{\mathbf k}
\label{eq2.1}
\end{equation}
where $\Psi_{\mathbf k}^n({\mathbf r})$ is a Bloch state,
$E_n({\mathbf k})$ is its eigenenergy and $E$ is complex in general.
For the periodic lattice both the total number of states and the
number of states within a given energy interval (in a band) is
infinite, but any numerical calculation can only provide states at a
finite number of ${\mathbf k}$-points, and for a finite number of
bands \cite{payne92}.  In view of these restrictions the approximation
of the spectral representation falls naturally into two parts -
approximating the Brillouin zone integral from a finite number of
$\mathbf k$-points, and approximating the infinite band sum.  It
should be noted that although it is well established that Green
function methods and the spectral representation can be defined within
a limited basis set \cite{williams82}, the embedding method cannot be
applied in this way since it relies on the properties of the Green
function in real space \cite{fisher88}.

\subsection{The Spectral function}
The simplest way to apply the defining equations for the embedding
potential is to expand the Green function in a set of basis functions
that are orthogonal over the embedding surface.  For the core
reconstruction considered here the embedding surface is a sphere
centred on the atomic site of interest, hence a natural set of basis
states are the spherical harmonics.  First the Kohn-Sham wavefunctions
are expanded on the surface $\mathcal S$,
\begin{equation}\Psi^{n}_{\mathbf k}({\mathbf r}_s)=
\sum_{\mathbf g} C_{\mathbf g}^{n}({\mathbf k}) e^{i({\mathbf
k}+{\mathbf g}).{\mathbf r}_s}= \sum_{L} \alpha^{(n)}_{L}({\mathbf k})
Y_{L}(\hat{\mathbf r})
\end{equation}
where $L=(lm)$, the combined index of the spherical harmonic $Y_L$,
and $C_{\mathbf g}^{n}({\mathbf k})$ are the expansion coefficients of
the eigenstates in the plane-wave representation.  The expansion
coefficients $\alpha^{(n)}_{L}({\mathbf k})$ can be found using the
identity \cite{morse53}
\begin{equation}
e^{i{\mathbf q}.{\mathbf r}}= 4\pi \sum_{L} i^l j_l(qr)
Y_{L}^*(\hat{\mathbf q}) Y_{L}(\hat{{\mathbf r}})
\end{equation}
hence
\begin{equation}
\alpha^{(n)}_{L}({\mathbf k})= 4\pi i^l \sum_{\mathbf g} j_l(|{\mathbf
k}+{\mathbf g}|r_s) Y^*_{L}(\widehat{{\mathbf k}+{\mathbf g}})
C_{\mathbf g}^{n}({\mathbf k}).
\end{equation}
From these a `spectral function' ${\mathcal
F}_{LL'}(E)=\frac{1}{\pi}\text{Im }[{\mathcal G}_{LL'}(E)]$ can be
defined, where ${\mathcal G}_{LL'}(E)$ are the expansion coefficients
for the Green function.  This spectral function is given by the
equation
\begin{equation}
{\mathcal F}_{LL'}(E)= \sum_n \int_{BZ} d^3{\mathbf k}\
\alpha^{(n)}_{L}({\mathbf k}) \alpha^{(n)*}_{L'}({\mathbf k})
\delta(E-E_n({\mathbf k}))
\label{eq3}
\end{equation}
and is related to the Green function by the convolution integral
\begin{equation}
{\mathcal G}_{LL'}(E)= \int_{-\infty}^{\infty} dE'\ \frac{{\mathcal
F}_{LL'}(E')}{E'-E}.
\label{eq4}
\end{equation}
In Eq.\ (\ref{eq3}) the integral on the right hand side reduces to the
surface integral
\begin{equation}
{\mathcal F}_{LL'}(E)=\sum_n \int_{E=E_n({\mathbf k})}
\frac{\alpha^{(n) }_{L }({\mathbf k}) \alpha^{(n)*}_{L'}({\mathbf k})
}{|\nabla_{\mathbf k}E|} dS
\label{eq5}
\end{equation}
 due to the delta function, hence evaluation of the spectral
representation reduces to the evaluation of a surface integral in
${\mathbf k}$-space and a singular convolution integral in energy
space.  The surface integral is carried out using the linear analytic
tetrahedron method
\cite{lehmann72,jepson71,macdonald79,rath75,kaprzyk86}, which results
in a spectral function that has the correct analytic structure in that
it is continuous in energy.  Only the irreducible wedge need be
sampled with the symmetry of the crystal used to complete the rest of
the integral \cite{trail98}.

In order to evaluate Eq.\ (\ref{eq4}) the singular integral itself
must be approximated from a sampling of the function at a finite
number of energy values.  Here the spectral function is interpolated
between consecutive energy values, and the contribution to the
integral from each of these ranges calculated.  We use a polynomial
interpolation, so the integrals can be evaluated analytically
\cite{davis67}.  This results in an approximation to the Green
function that can be evaluated for complex $E$ and which has the
analytic properties appropriate for a continuum of states; it is
essentially a generalisation of the approach used by Kuzmiak et al
\cite{kuzmiak91} to complex energies.  An alternative application of
the linear analytic tetrahedron method to the calculation of Green
functions is given Lambin and Vigneron \cite{lambin84} where Eq.\
(\ref{eq2.1}) is evaluated for each tetrahedron analytically, within
the linear interpolation scheme.  Although this approach is more
direct and introduces none of the errors due to approximating Eq.\
(\ref{eq4}), it requires the interpolation to be applied to all
tetrahedra, whereas the surface integral in Eq. (\ref{eq5}) requires
only a sub-set of tetrahedra to be considered for each energy.  In
addition to this the surface integral and convolution route allows a
greater flexibility in the degree of approximation applied, as
discussed below.

\subsection{Completing the Incomplete Set of States}
So far the evaluation of the Green function takes into account the
continuum nature of the band states, but for the spectral
representation to describe a real space Green function the set of
states used in Eq. (\ref{eq2.1}) must be complete.  We have found that
in order to obtain accurate Green functions using the spectral
representation, we must include the complete set of $n_{\mathbf g}$
states associated with the plane-wave basis set (for further
discussion see Trail\cite{trail98}).  $n_{\mathbf g}$ is the number of
plane-waves in the basis set, determined by the plane-wave energy
cut-off, and these states will be obtained by direct matrix
diagonalisation of the Kohn-Sham Hamiltonian.  This set of states is
still not complete in real space, hence we must consider the errors
introduced by not including the energy bands that are not available in
a finite plane-wave calculation.  In what follows we refer to the
spectral representation that includes only the $n_{\mathbf g}$ lowest
energy bands as the \emph{incomplete spectral representation}.  It is
important to realise that we mean the states are incomplete in real
space, they are, of course, complete in the sub-space spanned by the
plane-wave basis set.

It is not immediately apparent that this incompleteness will have a
significant effect, and it could be hoped that the `missing' high
energy states are so far above the energies of interest (ie at or
below the Fermi energy) that any error introduced by their absence
will be negligible.  This is only partly true, and the properties of
the error introduced by incompleteness are derived in appendix\
\ref{sec:appb} for a free electron gas.  Correcting for this
incompleteness not only speeds convergence with respect to $n_{\mathbf
g}$, it also ensures that the approximation of the Green function has
the correct analytic form.  We follow James and Woodley \cite{james96}
and approximate the high energy states `missing' from the incomplete
spectral representation by free electron states.  The plane-wave basis
set used to represent the lowest $n_{\mathbf g}$ bands is described by
$|{\mathbf g}|^2<E_{max}$, where ${\mathbf g}$ are reciprocal lattice
vectors and $E_{max}$ is the standard plane-wave cut-off energy.
Consequently, the free electron states required to `top up' the
incomplete spectral representation are those described in the reduced
zone scheme by $|{\mathbf g}|^2 \geq E_{max}$ and ${\mathbf k}$ in the
first Brillouin zone.

In order to calculate the required correction we calculate an
incomplete spectral representation for free electrons with exactly the
same basis as for the pseudopotential states, and subtract this from
the analytic free electron Green function.  This yields the
approximation
\begin{equation}
{\mathcal G}\approx {\mathcal G}^{pseudo}_{E_{max}}- {\mathcal
			G}^{free}_{E_{max}}+ {\mathcal
			G}^{free}_{\infty}
\end{equation}
where the first term on the right hand side is the incomplete spectral
representation of the pseudo-states, the second is the incomplete
spectral representation of free electron states and the final term the
complete spectral representation for free electron states (ie the
analytic free electron Green function).  In terms of the spectral
function, ${\mathcal F}$, and the convolution integral used to
transform this into the spectral representation, Eq. (\ref{eq4})
becomes
\begin{equation}
{\mathcal G}_{LL'}(E)=\int \frac{{\mathcal
F}^{pseudo}_{LL'}(E')-{\mathcal F}^{free}_{LL'}(E')}{E'-E} dE'
+{\mathcal G}^{\infty}_{LL'}(E)
\end{equation}
where ${\mathcal G}$ is now a Green function with the correct analytic
form, $E$ is complex and $E'$ is real.  ${\mathcal F}^{pseudo}$ and
${\mathcal F}^{free}$ are the spectral functions associated with the
pseudo-states and the free space states respectively, calculated with
$n_{\mathbf g}$ basis functions.  The last term, ${\mathcal
G}^{\infty}$, is the analytic free space Green function given by
\cite{morse53,abramowitz64}
\begin{equation}
{\mathcal G}^{\infty}_{LL'}=ikj_l(kr_s)h_l(kr_s) \delta_{ll'}
\end{equation}
where $j_l$ is the spherical Bessel function of the first kind, $h_l$
is the spherical Hankel function of the first kind and $E=k^2+V_0$,
where $V_0$ is the average potential within the unit cell.  In
practice the specific value of $V_0$ is not critical since the error
term varies only slowly with the Green function energy, as discussed
in appendix\ \ref{sec:appb}.  The normal derivative of the Green
function can be obtained by taking the radial derivative of the Green
function, and Eq.\ (\ref{eq2}) directly applied to give the embedding
potential (a factor of $\frac{1}{r_s^2}$ must be included to normalise
the spherical harmonics over the embedding sphere).  Eq. (\ref{eqa10})
was not used due to problems with small errors in the Green function
introducing anomalous poles at the bottom of the lowest band.

\section{Embedding Calculation}
\label{sec:reconstruction}
In this section we describe the all-electron calculation carried out
in the region near the nucleus, using the embedding potential
described above.  The normal density functional framework is used,
with the Kohn-Sham Hamiltonian extended by the addition of the
embedding terms.  Since these extra terms are functions of energy the
eigenvalue solution of the Hamiltonian is not simple, hence the charge
density is obtained directly from the Hamiltonian via the Green
function method as described by Williams et al \cite{williams82}.  The
method employed in the all-electron calculation is similar to that
used by other workers (eg Trioni et al \cite{trioni96}), but
generalised so as not to require any particular symmetry of the charge
density or self-consistent potential, and to include core electrons.

\subsection{The Embedded Hamiltonian}
Each basis function is a product of a radial function and a spherical
harmonic, with the radial part defined as an augmented plane-wave for
$r<s$ (region $B$) and a spherical Bessel function for $s<r<r_s$
(region $A$), where $s$ is a parameter of the calculation.  These are
chosen since Linearised Augmented Plane Waves (LAPW) orbitals describe
the all-electron valence states well near the nucleus, and in the
region nearer the embedding radius the spherical Bessel functions
provide the flexibility required to satisfy the boundary conditions.

The basis functions in region $B$ are found by solving the Dirac
equation using numerical integration in the spherical part of the
Kohn-Sham potential.  The method used is that described by Koelling
and Harmon \cite{koelling77}, where the Dirac equation is approximated
in the form of a Schr\"odinger equation which does not include
spin-orbit interaction but takes other relativistic effects into
account.  A solution is found at a fixed `pivot' energy, $E_p$, and
the resultant radial functions are denoted $u_{l}(r)$.  The energy
derivative of these functions are also obtained and orthogonalised to
the associated $u_{l}(r)$, and these orthogonalised energy derivatives
are denoted $\dot{u}_{l}(r)$ (see Krasovskii \cite{krasovskii97}, or
Takeda and K\"ubler \cite{takeda79}).  In $A$ the radial basis
functions used are spherical Bessel functions of the first kind,
defined as $j_l(g_{i}r)$ where $g_i=\frac{\pi i}{d}$.  The index $i$
is chosen to take integer values, to give a set of functions, and $d$
is a parameter of the calculation (larger than $r_s$ to allow
sufficient flexibility for the basis function to satisfy arbitrary
boundary conditions).  The radial parts of the basis functions,
$\chi_{il}(r)$ are therefore
\begin{equation}
\chi_{il}(r)=\left\{
\begin{array}{ll}
a_{il}u_{l}(r)+b_{il}\dot{u}_{l}(r) & 0 \leq r \leq s \\ j_l(g_ir) & s
< r \leq r_s \ .
\end{array}
\right.
\label{eq6}
\end{equation}
Parameters $a_{il}$ and $b_{il}$ are chosen to ensure $\chi_{il}$ is
continuous in amplitude and derivative at the $AB$ boundary.
Typically $r_s$ is chosen to be half the distance between nearest
neighbour atoms in the lattice, $s\sim 0.9r_s$ and $d\sim 2r_s$ to
give a good description of the valence electrons.  For $i$ the range
$1 \ldots 4$ is typical.

The embedded Hamiltonian matrix is then expanded in terms of these
basis functions, for a Kohn-Sham potential given by
\begin{equation}
V({\mathbf r})=\sum_{L} V_{L}(r) Y_{L}(\hat {\mathbf r})
\end{equation}
where $L=(lm)$, the index of the spherical harmonic.  We write
Hamiltonian matrix, ${\mathbf H}_{emb}$, as the sum of $3$ parts,
\begin{equation}
{\mathbf H}_{emb}={\mathbf H}^{A}+{\mathbf H}^{B}+{\mathbf \Sigma}
\end{equation}
with ${\mathbf H}^{A}$ the contribution from region $A$, ${\mathbf
H}^{B}$ the contribution from region $B$, and ${\mathbf \Sigma}$ the
contribution from the embedding terms at the surface ${\mathcal S}$.
Integrating over region $A$ results in
\begin{eqnarray}
H^A_{iL,jL'}= g_j^2 \int_s^{r_s}r^2 dr \left[ j_l(g_ir)j_l(g_jr)\right]
 \delta_{LL'} \nonumber\\ +\sum_{L''} S^L_{L'L''} \int_s^{r_s}r^2dr
 \left[ j_l(g_ir)V_{L''}(r)j_{l'}(g_jr) \right],
\end{eqnarray}
where $S^L_{L'L''}=\int Y_L^*Y_{L'}Y_{L''} d\Omega$ are Gaunt
coefficients.  For region $B$ we find a sum of spherical and
aspherical terms
\begin{eqnarray}
H^B_{iL,jL'}=&& \left[ a_{il}a_{jl} E_p \langle u_l \mid u_l \rangle
\right. \nonumber\\ && \left. + a_{il}b_{jl} \langle u_l \mid
\dot{u}_l \rangle + b_{il}b_{jl} E_p \langle \dot{u}_l \mid \dot{u}_l
\rangle \right]\delta_{LL'} \nonumber\\ &&+ \sum_{L''\neq0}
S^L_{L'L''} \int_0^{s}r^2dr \chi_{il}(r) V_{L''}(r) \chi_{jl'}(r)
\end{eqnarray}
where $E_p$ is the `pivot' energy at which $u_{l}(r)$ is calculated,
and the bra-ket's denote integration over region $B$ only.  The
integrals in the above expressions are carried out analytically or
numerically as appropriate.  Taking the normal derivative and
embedding potential terms in Eq.\ (\ref{eq1}) we obtain the
contribution from the embedding terms, ${\mathbf \Sigma}$, as
\begin{eqnarray}
\Sigma_{iL,jL'}=&&j_l(g_ir_s)\Gamma_{LL'}(E) j_{l'}(g_jr_s)
\nonumber\\ &&+ g_j j_{l}(g_ir_s) j'_{l'}(g_jr_s)\delta_{L,L'}.
\end{eqnarray}

In addition to the Hamiltonian matrix the overlap matrix of the basis
functions, $\mathbf O$, is also required.  This is given by
\begin{eqnarray}
O_{iL,jL'}= \delta_{LL'}\int_s^{r_s} j_l(g_ir) j_l(g_jr) r^2 dr
\nonumber\\ +\left[ a_{il}a_{jl} \langle u_l \mid u_l \rangle +
b_{il}b_{jl}\langle \dot{u}_{l} \mid \dot{u}_{l} \rangle
\right]\delta_{LL'}
\end{eqnarray}
where the first integral on the right hand side can be performed
analytically \cite{abramowitz64} and the second numerically.

\subsection{The Embedded Green function and Charge Density}  
The Green function of the embedded system is obtained directly by
matrix inversion,
\begin{equation}
{\mathbf G}(E)=\left( {\mathbf H}_{emb}(E)-E{\mathbf O} \right)^{-1},
\end{equation}
where ${\mathbf H}_{emb}$ is the embedded Hamiltonian described above.
To obtain the local density of states, $n({\mathbf r};E)$, we apply
the identity \cite{economou90}
\begin{equation}
n({\mathbf r};E)=\frac{1}{\pi}\text{Im } G({\mathbf r},{\mathbf r};E).
\end{equation}
By filling all states to the Fermi energy (which is obtained from the
original pseudopotential calculation using the linear analytic
tetrahedron method) we can obtain the valence charge density.
Unfortunately the number of points required to evaluate the charge
density by integrating the local density of states is fairly high, due
to its fine structure.  This can be avoided be taking advantage of
Cauchy's theorem \cite{williams82,morse53} and the analytic properties
of the Green function.  We choose a contour $C$ from some energy below
the lowest eigenvalue, following a half circle into the positive
complex plane and terminating on the real axis at the Fermi energy, so
the charge density expanded in spherical harmonics is given by
\begin{eqnarray}
\rho^{val}_{L''}(r)=&& \frac{1}{2\pi} \sum_{ijLL'}
\chi_{il}(r)\chi_{jl'}(r) \nonumber \\ &&\times \int_{C} \left(
G_{iLjL'}S^{L'}_{LL''}-G^*_{iLjL'}S^{L}_{L'L''} \right)
\label{eq6.1}
\end{eqnarray}
where the $G_{iLjL'}$ are evaluated at the complex energies on the
contour.  In this equation $S^{L}_{L'L''}$ is a Gaunt coefficient, and
is zero for $l'' > 2l$ or $2l'$, so a $l_{max} \times l_{max}$
Hamiltonian matrix will result in a charge density containing
components $l \leq 2l_{max}$.  It is due to the basis functions being
complex that this expression does not simply involve the imaginary
part of the Green function matrix.  As the imaginary part of the
energy increases the Green function becomes smoother and more
featureless, hence a more sparse sampling is sufficient to approximate
this integral accurately.  Gaussian integration \cite{press94} is used
to perform the integral along contour $C$, and converged integrals are
typically obtained for around $16$ points on the contour.

The core states could be obtained by finding the discrete states of
the embedded Hamiltonian, but this would be cumbersome (see appendix\
\ref{sec:appa}).  Core states are therefore obtained using the
Kohn-Sham potential within the embedding sphere, and a constant
potential outside of the sphere.  This gives accurate results as the
core states are strongly localised within the core region, so the
potential outside of the embedding sphere is largely irrelevant.  Only
the spherically symmetric part of the potential is used to obtain the
states, and the constant potential outside the sphere is taken to be
continuous with the spherical part.  This spherically symmetric
approximation has been used by many workers (eg Blaha et al
\cite{blaha88}; Methfessel and Frota-Pess\^{o} \cite{methfessel90}),
and can easily be extended to included aspherical effects
perturbatively as described by Ehmann and F\"ahnle \cite{ehmann97},
Sternheimer \cite{sternheimer86} and Lauer et al \cite{lauer79}.  To
obtain the core charge density a fully relativistic treatment is used,
solving the Dirac equation within the embedding region.

This approach allows the construction of core-states that are 
self-consistent within the embedded system, however the influence of core 
relaxation on the substrate system (and so on the embedding potential) 
is not considered.  Essentially we are assuming that the original 
pseudopotential approximation is valid, and that the valence 
pseudo-states are accurate outside of the core region. In order to 
go beyond this assumption it should be possible to obtain a new 
pseudopotential from the embedded core states and to iterate to full 
self-consistency, but this has not been implemented.  Similar 
considerations are discussed by Bl\"ochl \cite{blochl94}.  It should 
be noted that for the embedding method applied here the core states are 
truly localised, since we are embedding an all-electron atom at one 
site within a lattice of pseudo-atoms.

\subsection{Self Consistency}
The Kohn-Sham potential, $V^{KS}({\mathbf r})$, is calculated from the
total charge density via the Local Density Approximation (LDA), and
takes the form
\begin{equation}
V^{KS}_{L}(r)=V^{Nuc}_{L}(r)+V^{Hart}_{L}(r)+V^{XC}_{L}(r),
\label{eq7}
\end{equation}
a sum of the nuclear potential, the Hartree potential, and the
exchange-correlation potential respectively.  The nuclear potential is
assumed to be that of a point charge centred at the nucleus, and the
Hartree potential,
\begin{eqnarray}
V^{Hart}_{L}(r)=&&\frac{4\pi}{2l+1}\frac{1}{r^{l+1}} \int_0^r \left[
r'^{l} \rho_{L}(r') \right] r'^2 dr' \nonumber\\ && +
\frac{4\pi}{2l+1} r^l \int_r^{r_s} \left[ \frac{1}{r'^{l+1}}
\rho_{L}(r') \right] r'^2 dr' \nonumber\\ && + a_{L}r^l,
\label{eq8}
\end{eqnarray}
is obtained from Poisson's equation.  The constants $a_{L}$ are
defined by the boundary condition that the Kohn-Sham potential,
$V^{KS}_{L}(r)$, be equal to the self-consistent potential of the
original pseudopotential calculation on the embedding sphere.

The LDA is used to obtain the exchange-correlation potential from the
charge density using the Perdew and Zunger parameterisation
\cite{perdew81} of the results of Ceperly and Alder \cite{ceperly80}
(this was also used in the original pseudopotential calculation).
$V^{XC}({\mathbf r})$ is a non-linear function of the charge density,
so $V^{XC}_L(r)$ cannot be directly obtained from $\rho_L(r)$.
However, deviations from a spherical density are expected to be small,
so we expand the exchange-correlation potential as a Taylor series
\begin{equation}
V^{XC}(\rho)=V^{XC}(\rho_0)+ \left.  \frac{\partial V^{XC}}{\partial
\rho} \right|_{\rho_0} (\rho-\rho_0) + \cdots
\end{equation}
with $\rho_0$ the spherical part of the charge density.  A truncation
to linear order is sufficiently accurate (the quadratic terms are
negligible for the systems considered to date) and can be applied
directly.  Convergence to self-consistency is achieved, with
instabilities controlled and convergence speed increased by applying
Broyden mixing \cite{broyden65,johnson88}.

\section{Results}
\label{sec:results}
In order to test the effectiveness of the embedding method for core
reconstruction we examine aluminium in the FCC structure.  We begin by
giving an example of the calculated embedding potentials, and some
concrete evidence that the approximations made can give an accurate
embedding potential.  The initial self-consistent plane-wave
calculation is carried using the LDA, a plane-wave cut-off of $400$ eV
and 60 $\mathbf k$-points within the FCC irreducible wedge.  These
parameters are more than adequate to obtain effectively perfect
convergence, so we can attribute any errors in the reconstruction to
the embedding method.  A Kerker \cite{kerker80} pseudopotential is
used with a core radius $r_c=2.19\ au$, small enough for the core
regions not to overlap.  The resulting self-consistent potential is
then used to obtain the full set of eigenstates by matrix
diagonalisation for a plane-wave basis set $|{\mathbf g}|^2 <
E_{max}$, and at the $\mathbf k$-points required to carry out the
Brillouin zone integral using the linear analytic tetrahedron method.
These states are then used to construct the embedding potential, with
the embedded sphere taken as the `touching sphere' radius, $r_s=2.705\
au$.  The convergence behaviour of the embedding potential depends on
the number of plane-waves in the basis, $n_{\mathbf g}$, the number of
${\mathbf k}$-points in the irreducible wedge of the FCC Brillouin
zone, $n_{\mathbf k}$, and the spacing of energy points ($\Delta E$)
used to sample the spectral function, ${\mathcal F}$ (see
Eq. (\ref{eq4})).

\begin{figure*}[t]
\begin{center}
\includegraphics*{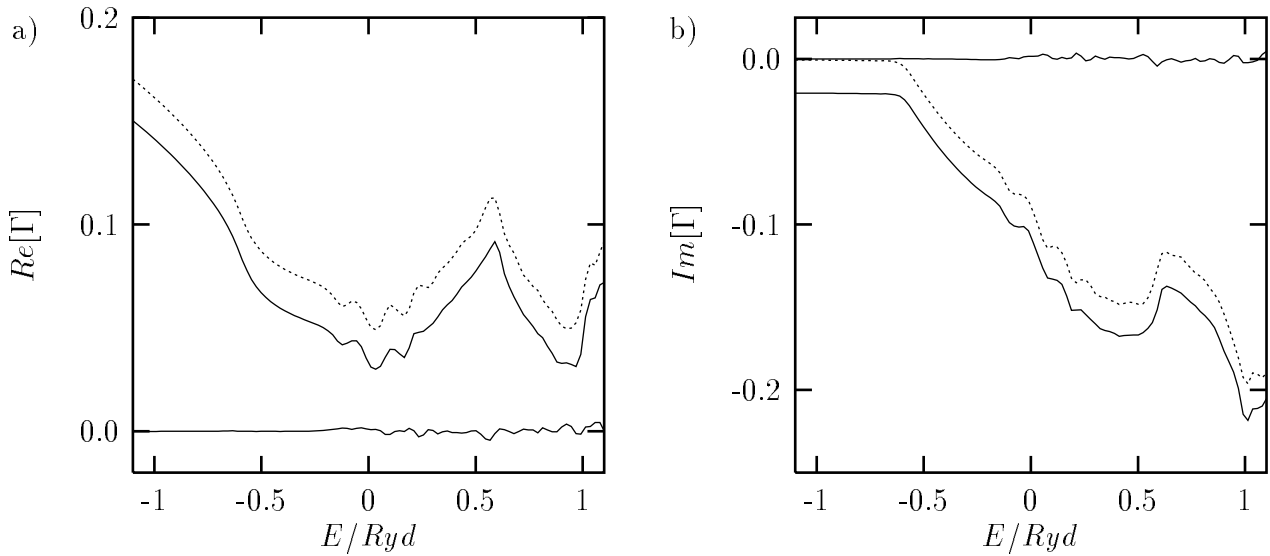}
\end{center}
\caption{(a) Real part, (b) imaginary part of aluminium embedding
potential matrix element, $\Gamma_{(10,10)}$, for $n_{\mathbf k}=240$
(solid line, for clarity shown offset by $-0.02$ for both real and 
imaginary parts) and $n_{\mathbf k}=505$ (dashed line).  The 
difference is also shown as the solid line near $\Gamma_{(10,10)}=0.0$.}
\label{fig2}
\end{figure*}

Fig.\ \ref{fig2} shows the $\Gamma_{(10,10)}$ matrix element for the
imaginary part of the energy equal to $0.1$ eV.  Results are shown for
$E_{max}=200$ eV and $\Delta E=0.3$ eV with linear interpolation of
the spectral function, and for both $240$ and $505$ $\mathbf k$-points
in the irreducible wedge of the Brillouin zone.  There is very little
discernable difference between the two results, $< 0.002$ for both the
real and imaginary parts, hence we consider $505$ $\mathbf k$-points
as effectively converged.  Using $89$ $\mathbf k$-points results in
larger errors ($< 0.01$), but was found to be adequate for performing
an accurate reconstruction, as discussed below.  The difference
between $E_{max}=200$ and $400$ eV is even smaller ($< 0.001$), and
comparing $\Delta E=0.3$ eV to $0.1$ eV again gives a difference of $<
0.001$.  The other matrix elements show a similar convergence
behaviour.  Embedding potentials for use in the reconstructions were
therefore calculated with $E_{max}=200$ eV, $\Delta E=0.3$ eV and
$240$ $\mathbf k$-points, for energies required on the contour of
integration (Eq. (\ref{eq6.1})).  Reconstructions were performed for
$l_{max}=6$, $d=4.0$, $4$ Bessel functions in the basis
(Eq. (\ref{eq6})), and 16 points along the contour of integration.
These provided the total charge density within the radius $r_s$, the
density of states within the embedded region, the self consistent
potential and the core eigenstates.

\subsection{Embedding with a Pseudopotential}
Before reconstructing the all-electron charge density, we carry out a
reconstruction in which no core states are included and the core and
nucleus are described by a pseudopotential, the same Kerker
pseudopotential used in the original plane-wave calculations.  This
provides a stringent test of the accuracy and reliability of the
entire embedding approach, since for a successful implementation the
resulting valence charge density should be the same as the original
plane-wave charge density throughout the embedding sphere.  The Fermi
energy used within the reconstruction is obtained from the original
plane-wave calculation, hence the total valence charge within the
embedding sphere will be different from the plane-wave calculation,
and the size of this difference provides a first indication of how
accurate the reconstruction is.  The original plane-wave calculation
has $2.298$ electrons within the embedding sphere, whereas the
reconstruction gives $2.300$ - an error of $0.08$ \%.  This close
agreement suggests a successful reconstruction, however this is a
fairly gross measure of success since it takes no account of the
structure of the electron states within the embedding sphere and
compares only the spherical part of the charge density.

\begin{table}
\begin{tabular}{llrr} \hline \hline
 & range & peak $(\times10^{-4}e\textrm{ }au^{-3})$ & $R(\%)$ \\
\colrule Pseudo & $r<r_s$ & -5.56 & 0.49 \\ Pseudo & $r_c<r<r_s$ &
-5.56 & 0.48 \\ All-electron & $r_c<r<r_s$ & -5.47 & 0.46 \\ \hline
\end{tabular}
\caption{Errors in charge density for each reconstruction.  Pseudo
refers to the reconstruction performed using a pseudopotential to
represent core states, and All-electron refers to the full
all-electron reconstruction.  The pseudopotential core radius is
$r_c=2.19\ au$, and the embedding sphere radius is $r_s=2.705\ au$.  }
\label{tab1}
\end{table}

A more demanding measure of the accuracy of the reconstruction is to
compare the charge density and/or self consistent potential of the
original pseudopotential calculation with the reconstruction.  The
error in the charge density is quantified as the peak error and the
$R$ factor \cite{press94} defined as
\begin{equation}
R=\int |\rho^{recon}({\mathbf r})-\rho^{pseud}({\mathbf r})|
d^3{\mathbf r} / \int |\rho^{pseud}({\mathbf r})| d^3{\mathbf r},
\end{equation}
and the errors over the whole embedding region are given in Table
\ref{tab1}.  We also give the error within the shell between the core
radius of the pseudopotential and the embedding sphere, which will
allow comparison with the all-electron results.

\begin{figure*}[t]
\begin{center}
\includegraphics*{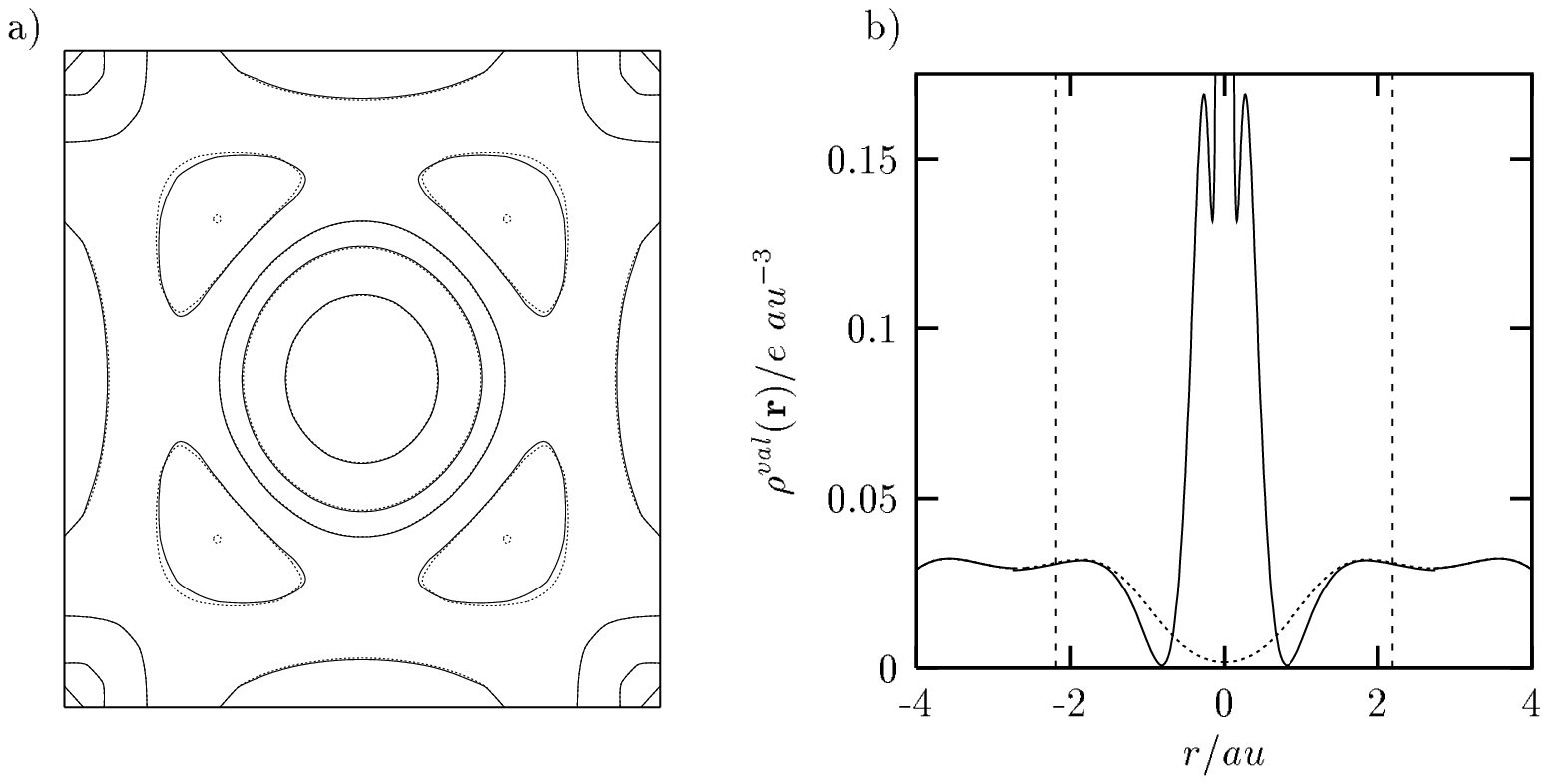}
\end{center}
\caption{ (a) Contours of constant valence charge density in the
\{100\} plane for the original plane-wave (dashed line) and
reconstructed pseudopotential (solid line) results.  The reconstructed
atom is at the centre and the square just encloses the embedding
sphere.  Contour levels are $0.01$, $0.02$, $0.025$, $0.030$ and
$0.032$ $e\textrm{ }au^{-3}$, chosen to emphasise the differences
between the original and reconstructed charge densities.  (b) Valence
charge density along a line in the [011] direction, with the
reconstructed atom at the origin.  Original plane-wave (dotted line)
and reconstructed all-electron results (solid line) are shown.  The
vertical dashed lines are at the core radius of the pseudopotential.}
\label{fig3}
\end{figure*}

Fig.\ \ref{fig3}a shows a contour plot of $\rho({\mathbf r})$, for
both the original plane-wave and reconstructed charge density.  These
are taken in the $\{100\}$ plane, with the embedded atom at the centre
and the square shown just enclosing the embedding sphere.  There is
excellent agreement between the original and reconstructed densities,
again demonstrating a successful reconstruction.

\subsection{The Reconstructed All-electron Charge Density}
The total valence charge within the embedding sphere is $2.299$
electrons for the all-electron reconstruction, which again agrees very
well with the original plane-wave result.  As above we also compare
the charge density of the original pseudopotential calculation with
the reconstruction.  For a successful reconstruction, the original
pseudopotential charge density should agree with the reconstructed
charge density for $r_c<r<r_s$.  Fig.\ \ref{fig3}b shows
$\rho^{val}({\mathbf r})$ along a line in the $[011]$ direction for
both the original plane-wave and all-electron reconstruction
calculations, with the reconstructed atom at the origin.  Agreement
between the two results is excellent in the region $r>r_c$.  The
errors between $r_c$ and $r_s$ are given in Table\ \ref{tab1}, and
these are similar to those of the pseudopotential reconstruction over
the same region.

As well as the charge density, errors in the self-consistent potential
have also been considered.  The reconstructed potential should agree
with the original (plane-wave) potential outside of the core radius.
The R-factor is smaller than for the charge density ($\sim 0.06\%$),
largely due to the fact that the reconstructed and original self
consistent potential are forced to agree at the embedding sphere.

\subsection{Original Pseudo and Reconstructed Density of States}
Another quantity whose accuracy is important is the Density of States
(DOS).  Since there is a $1:1$ correspondence between the all-electron
valence and pseudo-states, the eigenvalues are equal and the
pseudopotential is norm conserving, then a successful reconstruction
should yield an identical DOS to the original results.

\begin{figure}[t]
\begin{center}
\includegraphics*{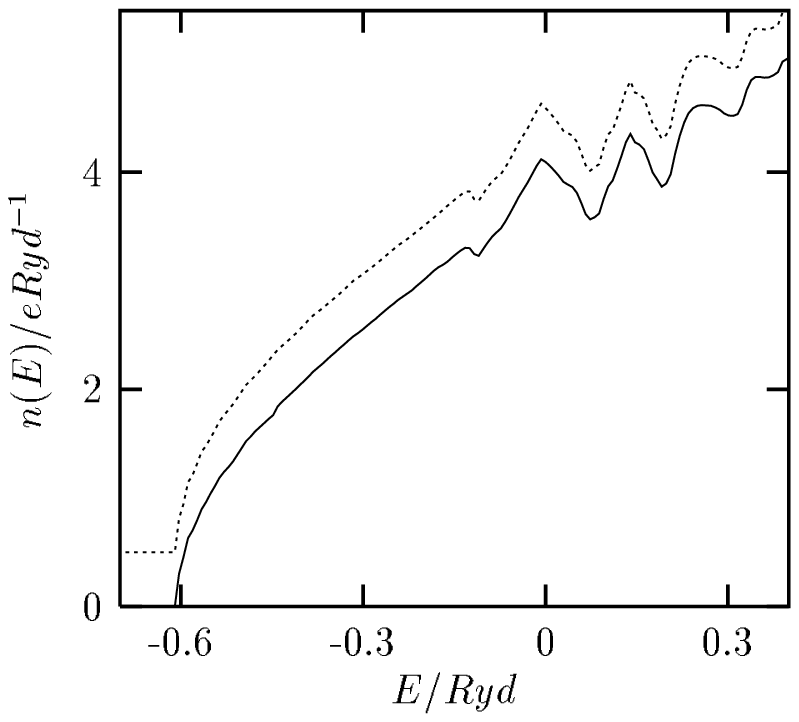}
\end{center}
\caption{Reconstructed all-electron (solid line) and original (dashed
line, for clarity shown offset by $0.5\mbox{ }e\mbox{ }Ryd^{-1}$) 
density of states within the embedding sphere.  Fermi energy is 
$0.207 Ryd$.}
\label{fig4}
\end{figure}

The DOS within $r_s$ is calculated from the plane-wave results by
applying the tetrahedron method to obtain the local density of states,
and integrating this over the volume contained by the embedding
sphere.  For the reconstructed DOS a self consistent calculation is
carried out, and the embedded Green function obtained along a contour
parallel to the real axis, with an imaginary energy of $0.1$ eV.  The
LDOS is taken from the imaginary part of this Green function, and
integrated over the volume of the embedding sphere, hence the
reconstructed DOS is smoothed by a Lorentzian of width $0.1$ eV, small
enough for the fine structure to be apparent.  In Fig.\ \ref{fig4} the
DOS within the embedding sphere is shown, for pseudo and all-electron
reconstructed states.  Agreement is excellent, with a maximum error of
$\sim 1$ \%.

\subsection{Convergence}
The results given above for the all-electron reconstruction were
reproduced for a number of different parameters of the embedding
potential and reconstruction calculation to assess the convergence of
the reconstructed results.  In what follows we define a charge density
as `well converged' when the associated $R$ value is less than $0.5$
\%.

First we consider the reconstruction calculation itself.  Convergence
was easily achieved with respect to the parameters of the
reconstruction calculation, that is, the number of basis functions in
Eq.\ (\ref{eq6}) and the number of points on the contour integral in
Eq.\ (\ref{eq6.1}).  The critical source of error in the
reconstruction was found to be the number of spherical harmonics used
in the basis, $l_{max}$.  A large enough value must be chosen for the
Green function within the reconstruction calculation to be converged,
and so to provide an accurate valence charge density.  For $l_{max}=6$
and $10$, $R=0.46$ \% and $0.18$ \% respectively, and although this
represents an improvement in the reconstruction the difference between
these two results very small.  It should also be noted that the error
introduced by a low $l_{max}$ is largely in the higher order
aspherical components of the charge density - the lowest order terms
present for aluminium ($L=(00),(4m)$) are very well converged for
$l_{max}=6$.

Second we consider the convergence of the reconstruction with respect
to the parameters of the embedding potential calculation.  These are
the number of $\mathbf{k}$-points at which the `band-structure' is
calculated, $n_{\mathbf k}$; the plane-wave cut-off energy used in the
calculations, $E_{max}$; and the interpolation of the spectral
function applied in Eq.\ (\ref{eq4}).  In general it was found that
the reconstructed charge density was well converged provided the
embedding potential was well converged, as discussed at the beginning
of this section.  The two main points of interest are the convergence
with respect to $n_{\mathbf k}$, and the sampling of the spectral
function used to carry out the convolution integral.

Reconstructions were performed for the smallest four $n_{\mathbf k}$
allowed within the linear analytic tetrahedron method, $n_{\mathbf
k}=20$, $89$, $240$ and $505$ in the irreducible wedge.  For these,
$R$ takes the values $2.30$, $0.47$, $0.44$ and $0.45$ \%
respectively, and there is no discernable difference between the
charge densities for $n_{\mathbf k}=89$ and $505$, suggesting that
convergence has been achieved for $89$ $\mathbf k$-points.  This
results is significant since the embedding potential itself was not
found to be particularly well converged for $89$ $\mathbf k$-points.
Turning to the convolution integral, it was found that by sampling at
energy intervals of $0.3$ eV below $5$ eV and $2.0$ eV above in Eq.\
(\ref{eq4}), the reconstructed results are indistinguishable from
those obtained by sampling at $0.1$ eV intervals throughout the energy
range - $R$ takes the value $0.57$ \% for the more sparse sampling and
$0.35$ \% for the finer sampling, and this difference is negligible.
This coarse sampling reduces the number of points required from $2878$
to $281$, significantly reducing the computational effort required to
construct the embedding potential.  With this sampling, if, for
example, the energy cut-off is increased from $200$ eV to $1000$ eV
the computational requirements are increased only by a factor of $\sim
3$, no matter how many bands are included in the calculation.

So far we have not justified using a linear interpolation of the
spectral function in order to carry out the convolution integral in
Sec.\ \ref{sec:green}.  When reconstruction results are compared for a
linear and cubic interpolation, and for a range of sampling intervals,
it is found that in the useful range (ie a fine enough sampling for
accurate results) cubic interpolation gives results only marginally
better, or even slightly worse than linear interpolation.  This is
probably due to the spectral function not being well approximated by a
polynomial, so a higher order interpolation can give worse results
than a low order method \cite{press94}.  It would be desirable to find
a better interpolating function, but the diverse analytic structure
due to van Hove singularities, and the requirement that an integral of
the form Eq. (\ref{eq4}) must be solved analytically, makes this a
non-trivial task.

\section{Conclusion}
A general method has been described for taking the results of a
pseudopotential calculation for a given system and obtaining the
correct all-electron charge density in the core region of a single
atom.  The reconstruction method provides correctly relaxed core
states (they are not frozen to the isolated atom core states) and is
in principle applicable to any system that can be analysed with
plane-wave pseudopotential methods.

The embedding potential method derived by Inglesfield
\cite{inglesfield81} is used, and a new analytic expression for the
embedding potential is given here.  Past applications of the embedding
potential method have generally been limited to models where the
embedding potential is that of free space, or an arbitrary model
embedding potential.  This is essentially due to the difficulty of
obtaining an accurate real space Green function for a realistic system
- the handful of cases where a Green function (and so embedding
potential) have been obtained from \emph{ab initio} calculations rely
on the properties of electron structure calculations that themselves
employ a Green function (for example see Thisjssen and
Inglesfield\cite{thijssen94}, Miller et al\cite{miller85},
Inglesfield\cite{inglesfield81b}, Crampin et al\cite{crampin92} or
Ishida\cite{ishida97}).  It has been found that in order to construct
an accurate real space Green function (and so embedding potential)
using the spectral representation a complete set of eigenstates must
be included, and here this has been taken into account by a linear
interpolation of the band structure in ${\mathbf k}$-space and
approximating the infinite number of high energy bands that are not
available as plane-wave states.  This results in an approximation that
will converge to the correct form and is accurate for a realisable
number of bands and ${\mathbf k}$-points.

Using this embedding method a localised all-electron atomic
calculation is carried out that does not require any assumption of
spherical symmetry in the potential necessary for previous solutions
to this problem, and in essence makes the same physical and
mathematical approximations as state of the art all-electron
density-functional methods such as the FLAPW method.  Results of tests
for aluminium are good - the original pseudopotential results can be
reconstructed with negligible error, and the all-electron results are
as accurate.  In addition to this the reconstruction accurately
reproduces the density of states of the original substrate system.

The major computational cost of performing a reconstruction lies not
in the reconstruction itself, but in the calculation of the full set
of states (ie band-structure) required to obtain an accurate real
space Green function.  For larger systems the computational cost of
obtaining the embedding potential from these states and performing the
reconstruction calculation are not expected to increase significantly,
but the cost of performing the matrix diagonalisation used to obtain
the full set of states could become prohibitive, scaling as
$n^3_{\mathbf g}$.  It would be advantageous to calculate the real
space Green function more efficiently, and it may be possible to
achieve this by applying iterative diagonalisation methods.  Provided
a full set of states of the self-consistent potential can be obtained,
we expect the method presented here to be applicable to any system.

In a future paper \cite{trail99} this method will be applied to the
all-electron reconstruction of the core region for bulk Si, with the
resulting charge density compared with accurate experimental
measurements of the structure factors and the results of FLAPW
calculations.

\begin{acknowledgements}
This work has been supported by United Kingdom
Engineering and Physical Sciences Research Council.  We thank
S. Crampin and J. E. Inglesfield for helpful discussions.
\end{acknowledgements}

\appendix

\section{Embedding Method}
\label{sec:appa}
Here we give a brief derivation of the embedding potential method of
Inglesfield \cite{inglesfield81}.  This is presented to shed some
light on the properties of the method, specifically the requirement
that the embedding potential be an analytic function defined even
where no states exist in the substrate system.  A slightly different
route is taken in the derivation of the embedding potential itself,
which yields the same expressions found
previously\cite{inglesfield81,fisher90}, together with a new
expression.

\subsection{The Embedded Hamiltonian}
Fig.\ \ref{fig1} shows the regions $I$, where the new embedded states
are obtained, and region $II$, the original substrate.  Although the
diagram shows the surface $\mathcal S$ as a sphere, this derivation
also applies to any other surface.  We proceed by finding the
expectation value of the energy of a trial function defined in $II$ as
the solution of the Schr\"odinger equation for the substrate at some
energy $\varepsilon$, $\psi({\bf r})$, and in $I$ as a trial function
${\phi({\bf r})}$.  This energy is then expressed in a form that is
dependent only on $\phi$ within region $I$ and the substrate Green
function on the surface, $\mathcal S$.  The variational principle is
then applied to obtain a Schr\"odinger equation which gives a solution
in region $I$ that correctly matches onto the solution in region $II$.
This is a normal Schr\"odinger equation defined in $I$ only and with
the addition of terms that are non-zero only at the surface $\mathcal
S$.

The equation for $\psi({\bf r})$ in $II$ is
\begin{equation}
(-\nabla^{2}+V({\bf r})-\varepsilon)\psi({\bf r})=0 \text{ }{\mathbf
r}\in II, \\
\end{equation}
hence the expectation value of the energy is given by
\begin{widetext}
\begin{equation}
E={ \frac { {\int_I d^3{\bf r} \phi^* H \phi } + {\varepsilon
\int_{II} d^3{\bf r} \psi^* \psi } + {\int_{\mathcal S} d^2{\bf r}_s
\left( \psi^*\left. \frac{\partial \phi}{\partial n_s}
\right|_{\bf{r}=\bf{ r}_s} - \phi^*\left. \frac{\partial
\psi}{\partial n_s} \right|_{\bf{r}=\bf{ r}_s} \right) } } { {\int_I
d^3{\bf r} \phi^* \phi } + {\int_{II} d^3{\bf r} \psi^* \psi }} }
\label{eqa1}
\end{equation}
\end{widetext}
where $H$ is the Hamiltonian and the spatial variable is suppressed.
The surface integral in the numerator is a consequence of the
discontinuity in the trial function at $\mathcal S$ and Green's
theorem, and can be interpreted as the contribution to the kinetic
energy from this discontinuity.  This expression has been used to
apply the variational method to trial functions with discontinuities
at a surface \cite{brownstein95}, but an additional condition
\begin{equation}
\phi(\bf{r_s})=\psi(\bf{r_s})
\label{eqa2}
\end{equation}
is chosen here.  It should be noted that no reduction in generality is
introduced by requiring this condition to be satisfied since the
solution in $II$ is not explicitly described at any point in the
derivation.

In order to carry out a calculation localised to region $I$ all
explicit dependence of $E$ on $\psi$ must be removed.  First we remove
the integral of $\psi$ over region $II$ using the expression
\cite{inglesfield81}
\begin{equation}
\int_{II} d^3{\bf r} |\psi({\bf r})|^2= \int_{S} d^2{\bf r}_s \left[
\psi^*({\bf r}_s) \frac{\partial }{\partial \varepsilon}
\frac{\partial \psi({\bf r}_s)}{\partial n_s} \right]
\label{eqa3}
\end{equation}
which is an aspherical generalisation of the similar expression that
describes the transferability of norm-conserving pseudopotentials
\cite{bachelet82}.  This leaves the energy $E$ dependent on $\psi$
only through the normal derivative at the surface.  We express this
normal derivative in terms of $\psi$ on the surface $\mathcal S$ as
\cite{inglesfield81}
\begin{equation}
\frac{\partial \psi({\bf r}_s)}{\partial n_s}=\int_{S} d^2{\bf r}_s'
\Gamma({\bf r}_s,{\bf r}_s';\varepsilon) \psi({\bf r}_s').
\label{eqa4}
\end{equation}
The function $\Gamma({\bf r}_s,{\bf r}_s';\varepsilon)$ is the
embedding potential, and by inserting Eq.\
(\ref{eqa2},\ref{eqa3},\ref{eqa4}) into Eq.\ (\ref{eqa1}) we obtain
the energy of the entire system expressed entirely in terms of the
trial function $\phi$ in region $I$.  The fact that an expression of
the form Eq.\ (\ref{eqa4}) exists is the heart of the embedding
method.

The condition that $E$ is stationary with respect to small changes
$\delta \phi$ gives the equation
\begin{eqnarray}
&\left(H+\delta({\bf r}-{\bf r}_s)\frac{\partial}{\partial
n_s}\right)\phi({\bf r})- &\nonumber\\ &{\delta({\bf r}-{\bf
r}_s)\int_{\mathcal S} d^2 {\bf r}_s' \left(\Gamma({\bf r}_s,{\bf r}
_s';\varepsilon) +(E-\varepsilon)\frac{\partial \Gamma({\bf r}_s,{\bf
r}_s';\varepsilon)}{\partial \varepsilon}\right)\phi({\bf r}_s')}
\nonumber\\ &{=E\phi({\bf r}) \ \ {\bf r} \in I }.
\label{eqa5}
\end{eqnarray}
for $\phi$.  This takes the form of a normal Schr\"odinger equation
with three additional terms - the derivative term and the two surface
integrals on the left hand side.  These extra terms take the form of a
non-local potential that acts only at the surface $\mathcal S$.

In deriving Eq.\ (\ref{eqa5}) the energy $E$ has not been minimised
with respect to variations in $\varepsilon$, hence $\varepsilon$ still
appears in the effective Schr\"odinger equation as a free parameter.
As a consequence of this the energy derivative term is present, a
first order correction to the energy at which the embedding potential
is evaluated.  This \emph{embedded Schr\"odinger equation} will give
the $\phi$ in region $I$ that is continuous with a solution in region
$II$ of energy $\varepsilon$ such that the combined trial function has
the lowest expectation value of energy.  If the expectation value of
the energy is further minimised with respect to $\varepsilon$, then
the combined trial function will be the eigenfunction of the complete
system of eigenenergy $E=\varepsilon$ \cite{trail98}.  In practice
this requires the eigenenergy of the solution to be known before the
equation can be directly solved, or iterative methods to be applied.

Within a continuum Eq. (\ref{eqa5}) takes a simpler form as the energy
of the state required can be chosen from the outset, and the first
order correction is zero.  In this case the embedded Hamiltonian for a
state of energy $E$ can be written
\begin{equation}
H_{emb}(E)= H_I+ \delta({\mathbf r}_{s}-{\mathbf r}) \left[
\frac{\partial}{\partial n_s} - \Gamma({\mathbf r}_{s},{\mathbf
r}'_{s};E) \right]
\label{eqa6}
\end{equation}
where $H_{emb}(E)$ is the embedded Hamiltonian that yields the states
with correct boundary conditions, and $H_I$ is the normal Hamiltonian
for region $I$.  The notation has been modified at this point such
that $\Gamma({\mathbf r}_{s},{\mathbf r}'_{s};E)$ acting on a function
denotes the integration over the surface $\mathcal S$, as in Eq.\
(\ref{eqa5}).

\subsection{Explicit Forms for the Embedding Potential}
In the above discussion the embedding potential, $\Gamma$, is defined
only implicitly with no explicit form given (or proof that any such
operator exists).  We now derive general forms for the embedding
potential in terms of the single particle Green function in region
$II$.  We start with the defining equation for the Green function in
region $II$,
\begin{equation}
(-\nabla^{2}_{r}+V({\bf r})-\varepsilon)G({\bf r},{\bf
r}';\varepsilon)= \delta( {\bf r}-{\bf r}') \ \ {\bf r},{\bf r}' \in
II.
\end{equation}
Multiplying this by $\psi$, and the Schr\"odinger equation for $\psi$
by $G({\bf r},{\bf r}';\varepsilon)$, subtracting the two, integrating
over region $II$ and applying Green's theorem gives
\cite{inglesfield81}
\begin{equation}
\psi=-{\mathcal G}.\frac{\partial\psi}{\partial n_s}+\frac{\partial
{\mathcal G}}{\partial n'_s}.\psi
\label{eqa7}
\end{equation}
where the surface integrals are represented as matrix multiplications.
This implies a representation of the spatial dependence in terms of a
set of basis functions which are orthonormal over the surface
$\mathcal S$; $\mathcal G$ denotes this representation.  It is
straightforward to show that for this representation the matrix
product corresponds to the integration over the surface $\mathcal{S}$.
The prime indicates that the normal derivative of the Green function 
is taken with respect to the second spatial variable.

The embedding potential required for the embedded Schr\"odinger
equation (Eq.\ (\ref{eqa6})) is the operator that gives the normal
derivative of a wavefunction on $\mathcal S$ in terms of its value on
$\mathcal S$, or
\begin{equation}
\frac{\partial\psi}{\partial n_s}= \Gamma.  \psi
\label{eqa8}
\end{equation}
where $\Gamma$ is the matrix representation of the embedding potential
operator.  Equations (\ref{eqa7}) and (\ref{eqa8}) lead immediately to
the expression
\begin{equation}
\Gamma=-{\mathcal G}^{-1}.\left( {\mathcal I}-\frac{\partial {\mathcal
G}}{\partial n'_s} \right).
\label{eqa9}
\end{equation}
This general expression for $\Gamma$ is the same as that originally
derived by Green function matching techniques
\cite{inglesfield81,garcia69,inglesfield71} and for von Neumann
boundary conditions reduces to the definition given by Inglesfield in
his original paper.

A second expression may be obtained by taking the normal derivative of
Eq.\ (\ref{eqa7}) with respect to the $1^{st}$ spatial variable, and
rearranging this to give
\begin{equation}
\Gamma=\left( {\mathcal I}+\frac{\partial {\mathcal G}}{\partial n_s}
\right )^{-1}.  \frac{\partial^2 {\mathcal G}}{\partial n_s \partial
n'_s},
\label{eqa10}
\end{equation}
which is an alternative and equally valid expression to Eq.\
(\ref{eqa9}) for the embedding potential in terms of a Green function
satisfying arbitrary boundary conditions on $\mathcal S$.  For
Dirichlet boundary conditions (${\mathcal G}=0$ for $\mathbf{r}$ or
$\mathbf{r'}$ $\in$ $\mathcal S$) this reduces to the form given by
Fisher \cite{fisher90}.

Finally, we note that at first glance the defining equation
(\ref{eqa8}) appears to suggest that the embedding potential can be
defined in terms of the eigenstates of the substrate system.  However,
this is not the case as Eq. (\ref{eqa8}) must be true for all energies
in order to apply the variational principle, even where there are no
substrate eigenstates.  In addition to this the eigenstates of the
embedded system are complex in the presence of a continuum of
substrate states, hence the embedding potential must be an analytic
function (for further discussion see Trail\cite{trail98}).

\section{Error in the Incomplete Spectral Representation for a Free 
Electron Gas}
\label{sec:appb}
In order to assess the properties of the error introduced by lack of
completeness of the spectral representation, the free electron case is
examined.  Since the high energy Bloch states should be essentially
free in character, it seems reasonable that this should give an
indication of the error introduced by applying the incomplete spectral
representation to a real system.  We calculate the error as the
contribution to the Green function from states of energy greater than
some maximum value, $E_{cut}=k_0^2$, using an extended zone scheme and
an expansion in spherical harmonics.

\begin{figure*}
\begin{center}
\includegraphics*{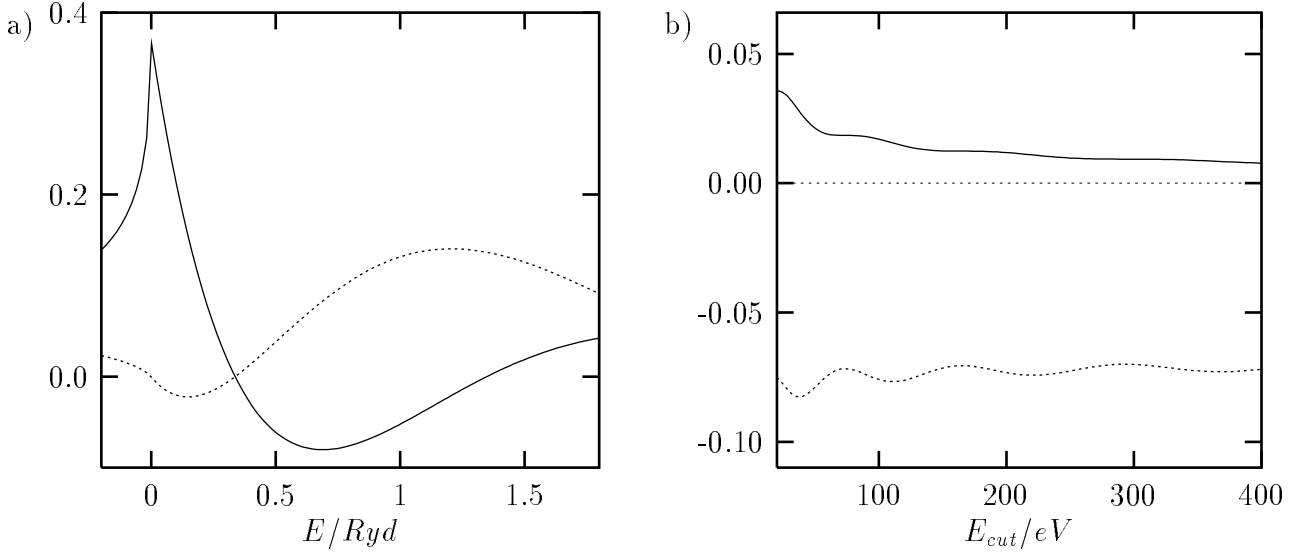}
\end{center}
\caption{(a) Real part of free electron Green function, ${\mathcal
G}_{(00,00)}$, (solid line) and its normal derivative (dashed line) as
a function of $E$.  (b) Error in ${\mathcal G}_{(00,00)}$ (solid line)
and its normal derivative (dashed line) at $E+i\varepsilon=0.0$ as a
function of $E_{cut}$.}
\label{fig5}
\end{figure*}

The free electron Green function is spherically symmetric, this is
also the case for the error, so both matrices are diagonal and
independent of the $m$ index.  In what follows only one index is
shown, the $m$ index is suppressed and the error term is denoted
$\xi_l(E,E_{cut})$.  The ${\mathcal F}$ matrix for free space is given
by the imaginary part of the Green function for a free electron, or
\begin{equation}
{\mathcal F}_{l}(E')=\frac{1}{\pi} k j_l(kr) j_l(kr')
\label{eqb1}
\end{equation}
where $E'=k^2$ and is real \cite{morse53}.  From this the error in the
Green function, $\xi_l(E,E_{cut})$, can be calculated by applying the
convolution integral (Eq. (\ref{eq4})) to the high energy states to
give
\begin{equation}
\xi_l(E+i\varepsilon,E_{cut})=\frac{2}{\pi} \int_{k_0}^{\infty}
k^2\frac{j_l^2(kr_s)}{k^2-(E+i\varepsilon)} dk,
\end{equation}
where the radial variables have been set equal to the embedding sphere
radius, $r_s$.  For $E_{cut} \gg |E+i\varepsilon|$ (true for all
$E+i\varepsilon$ we are interested in) a Taylor expansion can be used
to give the series
\begin{equation}
\xi_l(E+i\varepsilon,E_{cut})= \sum_{n} a_{l}^n (E+i\varepsilon)^n
\end{equation}
where
\begin{equation}
a_{l}^n(E_{cut})=\frac{2}{\pi} r_s^{2n-1} \int_{k_0 r_s}^{\infty}
\frac{j_l^2(x)}{x^{2n}} dx.
\end{equation}
These integrals can be carried out analytically by repeated
integration by parts, to give solutions of the form
\begin{equation}
a_{l}^n(E_{cut})=\frac{1}{\pi} \frac{1}{2n+1} \frac{1}{r_s^2}
\frac{1}{k_0^{2n+1}}+O\left( \frac{1}{k_0^{2n+2}} \right)
\end{equation}
where the second term is small for $|E+i\varepsilon| \ll E_{cut}$.

Since we also require the normal derivative of the Green function at
the surface (see Sec.\ \ref{sec:embedding}), the error in this is
derived in the same way.  We denote the coefficients of the normal
derivative series expansion by $b^n_l$, and find
\begin{eqnarray}
& b^n_l(E_{cut})=-\frac{1}{\pi} \frac{1}{2n+1} \frac{1}{r_s^3}
\left[1-\frac{2n+1}{2}(-1)^l \cos 2k_0r_s \right] \frac{1}{k_0^{2n+1}}
& \nonumber\\ & +O\left( \frac{1}{k_0^{2n+2}} \right)+ \delta_{n,0}
\left\{
\begin{array}{ll}
-1/2 r_s^2 & r' = r_s+0^{+} \\ \ 1/2 r_s^2 & r' = r_s+0^{-}
\end{array}
\right.  &
\label{eqb2}
\end{eqnarray}
where $r=r_s$ and the last term on the right hand side is a step
function whose value depends on which side we take the limit of $r'-r
\rightarrow 0$ in Eq.\ (\ref{eqb1}).  This step function is the source
of the cusp necessary for the correct analytic form of a Green
function, and will not be present in the incomplete spectral
representation.

For reasonable values of $E_{cut}$ we find that the errors in both
$\mathcal G$ and $\frac{\partial {\mathcal G}}{\partial n_s'}$ are
nearly constant in $E+i\varepsilon$, so for convenience we examine the
errors at $E+i\varepsilon = 0$ and $l=0$.  Fig.\ \ref{fig5}a shows the
real part of the analytic Green function and its normal derivative, as
a function of real $E$ and for $l=0$.  Alongside this, in Fig.\
\ref{fig5}b, the error in the Green function and its normal derivative
are shown as a function of $E_{cut}$, and at $E+i\varepsilon=0$.
These are obtained by evaluating $a_0^0$ and $b^0_0$ analytically.  It
is apparent that the errors are significant and that convergence with
respect to the energy cut-off is slow.  For the normal derivative the
error does not converge to zero with increasing $E_{cut}$ due to the
step function in Eq.\ (\ref{eqb2}).  For $l>0$ the errors show a
similar behaviour.



\begin{thebibliography}{}
\bibitem{payne92} M. C. Payne, M. P. Teter, D. C. Allan, T. A. Arias
and J. D. Joannopoulos, Rev.\ Mod.\ Phys. {\bf 64}, 4 1045 (1992).
\bibitem{hill98} G. J. Hill, J. M. Keartland, M. J. R. Hoch and H.
Haas, Phys.\ Rev.\ B {\bf 58}, 13614 (1998).
\bibitem{blochl94} P. E. Bl\"ochl, Phys.\ Rev.\ B {\bf 50}, 17953
(1994).
\bibitem{turzhevsky94} S. A. Turzhevsky, D. L. Novikov, V. A. Gubanov
and A. J. Freeman, Phys.\ Rev.\ B {\bf 50}, 3200 (1994).
\bibitem{huhne98} T. Huhne, C. Zecha, H. Ebert, P. H. Dederichs and
R. Zeller, Phys.\ Rev.\ B {\bf 58}, 10236 (1998).
\bibitem{goringe97} C. M. Goringe, D. R. Bowler and E. Hern\'andez,
Rep.\ Prog.\ Phys. {\bf 60}, 1447 (1997).
\bibitem{gardner86} J. R. Gardner and N. A. W. Holzwarth, Phys.\
Rev. B {\bf 33}, 10 7139 (1986).
\bibitem{vackar94} J. Vack\'a\v{r} and A. \v{S}im\.unek, J.Phys.:\
Condens.\ Matter {\bf 6}, 3025 (1994).
\bibitem{kuzmiak91} V. Kuzmiak, J.Zavadil and K. \v{Z}\v{d}\'ansk\'y,
Phys.\ Stat.\ Sol.(b) {\bf 168}, 547 (1991).
\bibitem{meyer95} B. Meyer, K. Hummler, C. Els\"asser and M. F\"ahnle,
J.Phys.: Condens.\ Matter {\bf 7}, 9201 (1995).
\bibitem{inglesfield81} J. E. Inglesfield, J.Phys.\ C {\bf 14}, 3795
(1981).
\bibitem{williams82} A. R.  Williams, P. J. Feibelman and N. D.  Lang,
Phys.\ Rev.\ B {\bf 26}, 10 5433 (1982).
\bibitem{fisher88} A. J. Fisher J.Phys.\ C {\bf 21}, 3229 (1988).
\bibitem{morse53} P. M.  Morse and H. Feshbach, \emph{Methods of
Theoretical Physics} (Plenum Press, New York, 1953).
\bibitem{lehmann72} G. Lehman and M. Taut, Phys.\ Stat.\ Sol.(b) {\bf
54}, 469 (1972).
\bibitem{jepson71} O. Jepson and O. K. Anderson, Solid\ State\
Commun. {\bf 9}, 1763 (1971).
\bibitem{macdonald79} A. H. MacDonald, S. H. Vosko and P.T. Coleridge,
J.Phys.\ C {\bf 12}, 2991 (1978).
\bibitem{rath75} J. Rath and A. J. Freeman, Phys.\ Rev.\ B {\bf 11}, 6
2109 (1975).
\bibitem{kaprzyk86} S. Kaprzyk and P. E. Mijnarends, J.Phys.\ C {\bf
19}, 1283 (1986).
\bibitem{trail98} J. R. Trail, \emph{Core Reconstruction in
Pseudopotential Calculations}, Phd Thesis, University of Bath, 1998.
\bibitem{davis67} P. J. Davis and P. Rabinowitz, \emph{Numerical
Integration} (Academic Press, New York, 1967).
\bibitem{lambin84} Ph. Lambin and J. P. Vigneron, Phys.\ Rev.\ B {\bf
29}, 6 3430 (1984).
\bibitem{james96} R. James and S. M. Woodley, Solid\ State\
Commun. {\bf 97}, 11 935 (1996).
\bibitem{abramowitz64} M. Abramowitz and I. A. Stegun, \emph{Handbook
of Mathematical Functions} (U.S. Government Printing Office,
Washington D.C., 1964).
\bibitem{trioni96} M. I. Trioni, G. P. Brivio, S. Crampin and
J. E. Inglesfield, Phys.\ Rev.\ B {\bf 53}, 12 8052 (1996).
\bibitem{koelling77} D. D. Koelling and B. N. Harmon, J.Phys.\ C {\bf
10}, 3107 (1977).
\bibitem{krasovskii97} E. E. Krasovskii, Phys.\ Rev.\ B {\bf 56}, 20
12866 (1997).
\bibitem{takeda79} T. Takeda and J. K\"ubler, J.Phys.\ F {\bf 9}, 4
661 (1979).
\bibitem{economou90} E. N. Economou, \emph{Green's Functions in
Quantum Physics} (Springer-Verlag, Berlin, 1990)
\bibitem{press94} W. H. Press, S. A. Teukolsky, W. T. Vetterling and
B. P. Flannery, \emph{Numerical Recipes} (Cambridge University Press,
Cambridge, 1994).
\bibitem{blaha88} P. Blaha, K. Schwarz and P. H. Dederichs, Phys.\
Rev.\ B {\bf37}, 2792 (1988).
\bibitem{methfessel90} M. S. Methfessel and S. Frota-Pess\^{o},
J.Phys.:\ Condens.\ Matter {\bf 2}, 149 (1990).
\bibitem{ehmann97} J. Ehmann and M. F\"ahnle, Phys.\ Rev.\ B {\bf 55},
12 7478 (1997).
\bibitem{sternheimer86} R. M. Sternheimer, Z. Naturforsch {\bf 41a},
24 (1986).
\bibitem{lauer79} S. Lauer, V. R. Marathe and A. Trautwein, Phys.\
Rev.\ A {\bf 19}, 5 1852 (1979).
\bibitem{perdew81} J. P. Perdew and A. Zunger, Phys.\ Rev.\ B {\bf
23}, 10 5048 (1981).
\bibitem{ceperly80} D. M. Ceperly and B. J. Alder, Phys.\ Rev.\
Lett. {\bf 45}, 566 (1980).
\bibitem{broyden65} C. G. Broyden, Math.\ Comput. {\bf 19}, 577
(1965).
\bibitem{johnson88} D. D. Johnson, Phys.\ Rev.\ B {\bf 38}, 18 12807
(1988).
\bibitem{kerker80} G. P. Kerker, J.Phys.\ C {\bf 13}, L189 (1980).
\bibitem{thijssen94} J. M. Thisjssen and J. E. Inglesfield,
Europhys. Lett. {\bf 23}, 65 (1994).
\bibitem{miller85} N. C. Miller, P. M. Lee and J. E. Inglesfield,
Phil.\ Mag. {\bf 51}, 209 (1985).
\bibitem{inglesfield81b}J. E. Inglesfield, J.\ Phys.\ F {\bf 11}, L287
(1981).
\bibitem{crampin92} S. Crampin, J. B. A. N. Van Hoof, M. Nekovee,
J. E. Inglesfield, J. Phys.: Condens. Matter {\bf 4} 1475 (1992).
\bibitem{ishida97} H. Ishida, Surf. Science {\bf 388}, 71 (1997).
\bibitem{trail99} J. R. Trail and D. M. Bird, Phys.\ Rev.\ B 
{\bf 60}, 7875 (1999).
\bibitem{fisher90} A. J. Fisher, J.Phys.:\ Condens.\ Matter {\bf 2},
6079 (1990).
\bibitem{brownstein95} K. R. Brownstein, J.Math.\ Phys. {\bf 36}, 1 76
(1995).
\bibitem{bachelet82} G. B. Bachelet, D. R. Hamann and M. Schl\"uter,
Phys.\ Rev.\ B {\bf 26}, 8 4199 (1982).
\bibitem{garcia69} F. Garcia-Moliner and J. Rubio, J.Phys.\ C {\bf 2},
1789 (1969).
\bibitem{inglesfield71} J. E. Inglesfield, J.Phys.\ C {\bf 4}, L14
(1971).
\end{thebibliography}
\end{document}